\documentclass[%
 aip,
rsi,%
 amsmath,amssymb,
 reprint,%
longbibliography,
]{revtex4-1}

\usepackage{color}
\usepackage{graphicx}
\usepackage{dcolumn}
\usepackage{bm}

\begin{document}

\preprint{AIP/123-QED}

\title[Homogeneous enhancement of electric near-field]{Homogeneous  enhancement of electric  near-field in all-dielectric metasurfaces composed of cluster-based unit cells} 

\author{Anton~S.~Kupriianov} 
\affiliation{College of Physics, Jilin University, 2699 Qianjin St., Changchun 130012, China}
\author{Kateryna L. Domina} 
\affiliation{School of Radio Physics, V. N. Karazin Kharkiv National University, 4, Svobody Sq., Kharkiv 61022, Ukraine}
\author{Vyacheslav~V.~Khardikov}
\affiliation{School of Radio Physics, V. N. Karazin Kharkiv National University, 4, Svobody Sq., Kharkiv 61022, Ukraine}
\affiliation{Institute of Radio Astronomy of National Academy of Sciences of Ukraine, 4 Mystetstv St., Kharkiv 61002, Ukraine}
\author{Andrey B. Evlyukhin}
\affiliation{Institute of Quantum Optics, Leibniz Universit\"at Hannover, 30167 Hannover, Germany}
\author{Vladimir~R.~Tuz}
 \email{tvr@rian.kharkov.ua and tvr@jlu.edu.cn}
\affiliation{Institute of Radio Astronomy of National Academy of Sciences of Ukraine, 4 Mystetstv St., Kharkiv 61002, Ukraine}
\affiliation{International Center of Future Science, State Key Laboratory of Integrated Optoelectronics, College of Electronic Science and Engineering, Jilin University, 2699 Qianjin St., Changchun 130012, China}

\date{\today}

\begin{abstract}
In order to construct a dielectric analog of spaser, we study theoretically and experimentally several configurations of cluster-based unit cells for an all-dielectric metasurface characterized by resonant conditions of the trapped mode excitation. Excitation of the trapped mode is realized either by specific displacement of particles in the cluster, or by perturbation of the equidistantly spaced particles by off-centered round holes or coaxial-sector notches. It turns out that the latter approach is more advantageous for enhancement of electric near-field with homogeneous distribution in-plane of the structure and its strong localization outside the high-refractive-index dielectric particles. This feature opens prospects for realization of highly desirable subwavelength flat lasing structures based on strong near-field interaction with substances exhibiting pronounced nonlinear characteristics and properties of gain media.
\end{abstract}



\maketitle

Considerable interest in the study of metamaterials is due to the prospects of their use in practical devices.\cite{Zheludev_NatMater_2012} Metamaterials can be a suitable platform for many optical systems, such as sensors\cite{Chen_Sensors_2012} and perfect absorbers.\cite{Radi_PhysRevApplied_2015} They allow one to enhance quantum dots luminescence,\cite{Tanaka_PhysRevLett_2010, Staude_ACSPhotonics_2015} realize optical switching \cite{tuz_PhysRevB_2010, Tuz_JOpt_2012} and other related operations \cite{Zheludev_LightSciAppl_2015} when combined with optically active and nonlinear substances.\cite{Lapine_RevModPhys_2014, Khardikov_Springer_2016} In the latter case, thin planar metamaterials (metasurfaces) are of special interest, due to their higher workability.\cite{Holloway_IEEE_2012, Glybovski_PhysRep_2016} 

In particular, it is proposed to combine metasurfaces with optically active materials to obtain parametric gain systems and develop amplifying or lasing devices \cite{Zheludev_NatPhotonics_2008} (e.g. spaser -- Surface Plasmon Amplification by Stimulated Emission of Radiation \cite{Bergman_PhysRevLett_2003, Premaratne_AdvOptPhoton_2017}). In a metasurface-based spaser a regular array of subwavelength metallic resonators is supported by a slab of gain medium containing quantum dots. A special type of symmetry-broken resonators is chosen to ensure excitation of a high-quality-factor (high-$Q$) trapped mode with reduced radiative losses.\cite{Fedotov_PhysRevLett_2007, khardikov_JOpt_2010} The collective plasmonic oscillations in such resonators lead to the emission of spatially and temporarily coherent light in the direction normal to the metasurface array. The spaser system is very thin and compact and benefits from the strong electric field localization near the surface associated with plasmons. Nevertheless, although the concept of the metasurface-based spaser is well developed, its practical implementation is difficult due to requirement of high pumping power, which adversely affects the system. It arises from excessive heat losses inherent in plasmonic nanostructures in infrared and visible parts of spectrum. Moreover, asymmetric plasmonic particles composing the metasurface typically have a quite complicated form, so it is difficult to fabricate them on the nanoscale.

All-dielectric metasurfaces can overcome above-mentioned drawbacks of plasmonic structures while being simple in manufacturing.\cite{zywietz2014laser, Kuznetsov_Science_2016, Kruk_ACSPhotonics_2017} The resonant behavior of light in high-refractive-index (high-$n$) dielectric nanoparticles\cite{Evlyukhin_PhysRevB_2010, evlyukhin2011multipole} makes it possible to reproduce many subwavelength effects demonstrated in plasmonic systems due to the electric field localization, but without much losses and energy dissipation into heat. In addition, the coexistence of strong electric and magnetic multipolar resonances, as well as their interference and corresponding  enhancement of near-fields in dielectric nanoparticles\cite{evlyukhin2012demonstration} bring much novel functionality to simple geometries, especially for the nonlinear regimes\cite{Smirnova_Optica_2016, Tong_OptExpress_2016, Liu_NatPhys_2018} and metadevices widespread applications.\cite{Chong_NanoLett_2015} However, since the electric near-field is mainly localized inside the high-$n$ particles, it interacts only with a small portion of the surrounding gain medium, which limits the optical output of the overall lasing system. This fact impairs advantageous use of such metasurfaces in construction of a dielectric analog of spaser.

Combining dielectric particles into clusters makes it possible to overcome this obstacle. For instance, in an all-dielectric metasurface composed of bars of different length a trapped mode can be excited.\cite{Khardikov_JOpt_2012, Zhang_OptExpress_2013} When resonant conditions are satisfied for the trapped mode, large electric near-field enhancement and localization in the surrounding medium appears inside the nano gaps introduced at the centre of bars.\cite{Zhang_OptExpress_2014} Moreover, it was recently reported \cite{Tuz_ACSPhotonics_2018, Tuz_OptExpress_2018, Yu_JApplPhys_2019, Sayanskiy_PhysRevB_2019, Kupriianov_PhysRevApplied_2019} that a proper choice of asymmetric dielectric particles and their ordering into clusters provides advanced flexibility in obtaining the desired near-field configuration, together with significant field localization caused by the trapped mode excitation. 

In development of the concept of cluster-based metasurfaces, in the present Letter we propose and characterize several configurations of an all-dielectric metasurface whose array sustains resonant conditions of the trapped mode. We demonstrate that under these conditions the resonant electric near-field appears to be homogeneously distributed in-plane of the structure and is strongly localized outside the high-$n$ dielectric particles. Our finding is confirmed by both numerical simulations and proof-of-principle microwave measurements of the near-fields in the actual metasurface prototypes.

\begin{figure}[t!]
\centering
\includegraphics[width=1.0\linewidth]{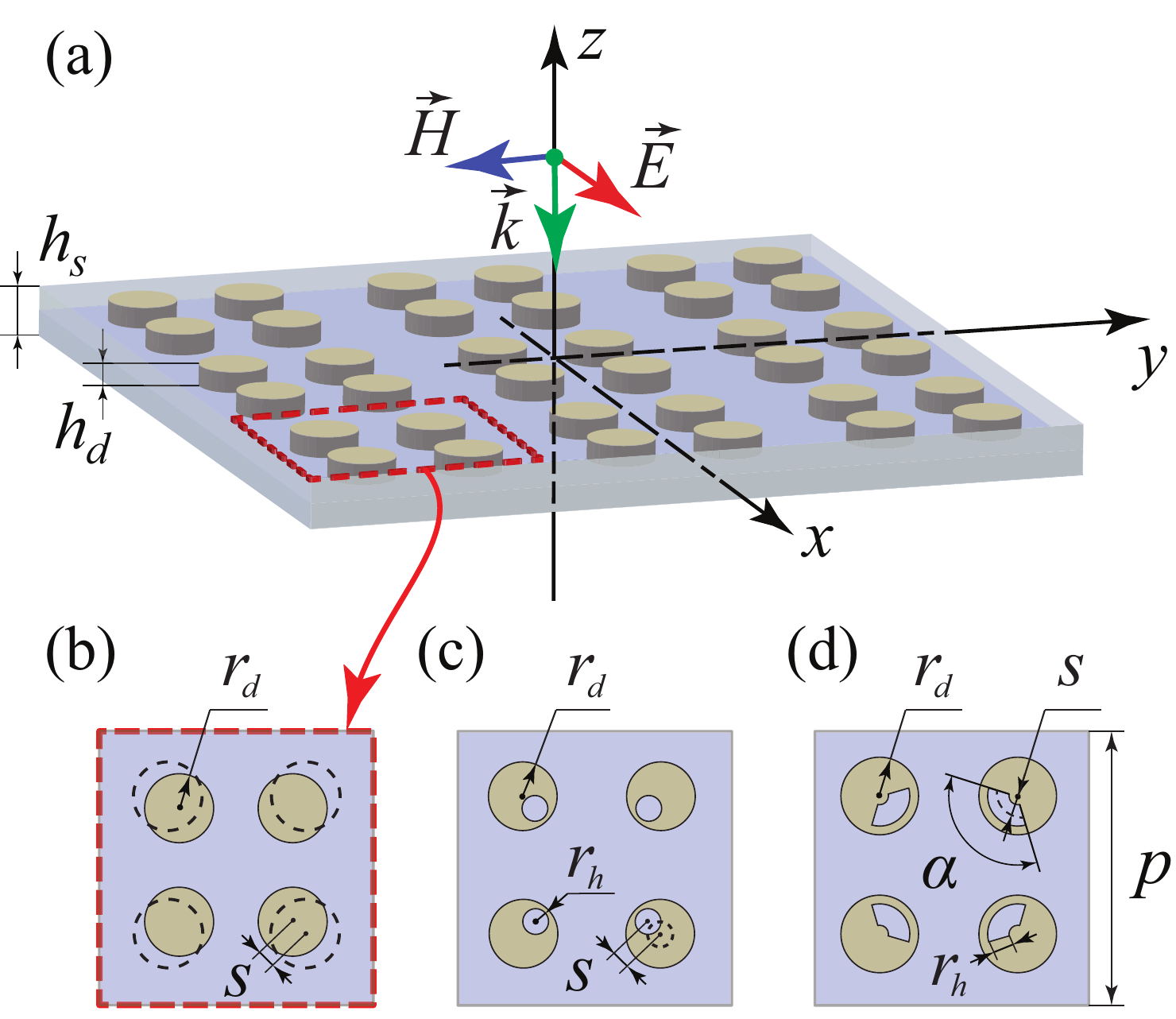}
\caption{(a) Schematic view of an all-dielectric metasurface whose cluster-based unit cell is composed of (b) solid disks shifted to the center of unit cell along its diagonals, (c) disks with off-centered round holes, and (d) disks with coaxial-sector notches.}
\label{fig:sample}
\end{figure}

In what follows, we perform a numerical and experimental study of resonant characteristics of an all-dielectric metasurface whose unit cell is composed of four cylindrical resonators (disks) [see Fig.~\ref{fig:sample}(a)]. The size of a square unit cell is $p$. Disks are made of a nonmagnetic dielectric material with permittivity $\varepsilon_d$. The radius and thickness of the disks are $r_d$ and $h_d$, respectively. The disks are immersed into a dielectric substrate (host) with permittivity $\varepsilon_s$ and thickness $h_s$.

It was theoretically shown\cite{Evlyukhin_PhysRevB_2010, vandeGroep_OptExpress_2013, babicheva2019analytical} and experimentally confirmed \cite{Liu_OptExpress_2017} that a metasurface consisting of an array of equidistantly spaced dielectric particles demonstrate a resonant response arising from excitation of electric and magnetic multipole moments of individual particles. If such particles (e.g. dielectric disks) are arranged into clusters, the inter-particle coupling leads to a complex collective behavior of modes which is different from that of individual particles.\cite{Tuz_ACSPhotonics_2018,terekhov2019multipole} Moreover, specific perturbations of the particles can result in the appearance of additional ultra high-$Q$ resonances in the overall metasurface response. These resonances are related to the collective trapped mode excitation.\cite{Tuz_OptExpress_2018, Yu_JApplPhys_2019, Sayanskiy_PhysRevB_2019}

Therefore, in order to differentiate the features of the cluster-based configuration from those influenced by the particles perturbations, further we consider three particular designs of the metasurface. The first design is realized by displacing the non-perturbed disks toward the unit cell center by a distance $s$ along the cell's diagonals [Fig.~\ref{fig:sample}(b)]. In the second design, the disks of the array are equally spaced, whereas they are perturbed by an eccentric through hole. The hole radius is indicated as $r_h$, the distance from the center of the disk to the center of the hole is $s$. Within the unit cell all holes are oriented inward the center of the cluster, as shown in Fig.~\ref{fig:sample}(c). In the third design, equally spaced disks are perturbed by a coaxial-sector notch [Fig.~\ref{fig:sample}(d), see also discussion of benefits of such perturbation approach in Refs.~\onlinecite{Tuz_OptExpress_2018, Sayanskiy_PhysRevB_2019}]. The coaxial-sector notch is characterized by the radius of the middle line $s$, width $r_h$, and opening angle $\alpha$. The notches of the disks are oriented inward the center of the unit cell. For all designs we define the dimensionless asymmetry parameter $\theta$ of the metasurface. The parameter $\theta$ varies in the range $[0-1]$ and is associated with the displacement of the disks toward the cluster center $[\theta=2s/(p-2r_d)]$, increase in the radius of the circular hole $(\theta=2r_h/r_d)$, and the opening angle of the notch $[\theta=\sin(\alpha/2)]$ for the first, second, and third design, respectively.

The cluster forming metasurface corresponds to the group $C_{4v}$ for all chosen designs.\cite{Yu_JApplPhys_2019} This group has the four-fold symmetry axis with respect to rotation around the $z$ axis. For the metasurfaces whose unit cell symmetry belongs to the rotational groups $C_{4v}$, there is polarization independence of the structure under normal incidence conditions.\cite{Sayanskiy_PhysRevB_2019} Therefore, in what follows we consider irradiation of the metasurface by a normally incident plane wave $(\vec k = \{0,0, -k_z\})$, where the electric field vector is directed along the $x$ axis $(\vec E=\{E_x,0,0\})$. 

\begin{figure}[t!]
\centering
\includegraphics[width=1\linewidth]{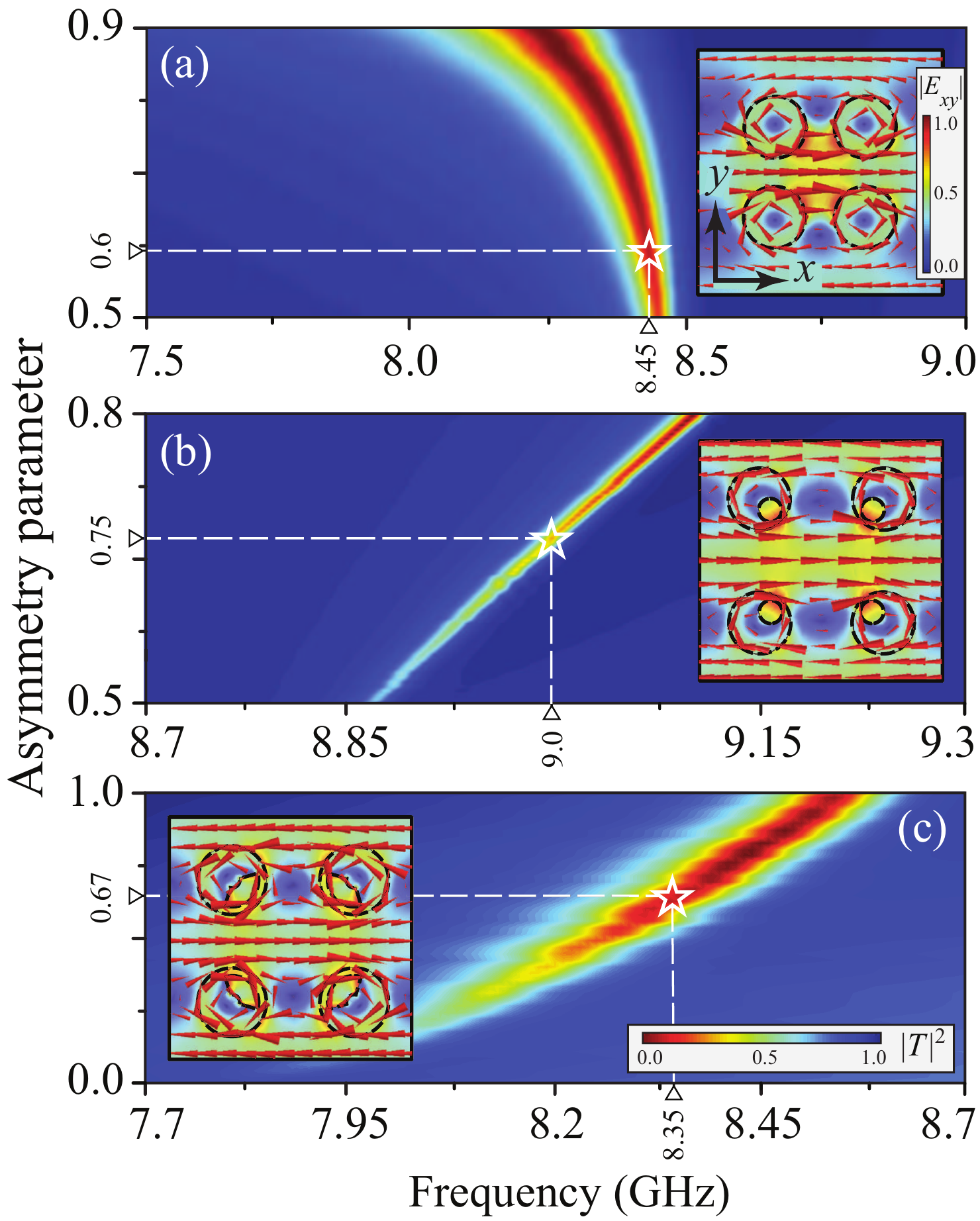}
\caption{Transmission versus frequency and asymmetry parameter for the idealized (loss-less) all-dielectric metasurface whose cluster-based unit cell is composed of (a) non-perturbed disks ($r_d=4.0$ mm), (b) disks perturbed by a round off-centered hole ($r_d=4.0$ mm, $s=2.0$ mm), and (c) disks perturbed by a coaxial-sector notch ($r_d=4.5$ mm, $r_h=2.0$ mm, $s=2.0$ mm). Insets present magnitude and vectorial patterns of the electric near-field calculated in the cross section along the midline of the resonators ($z=0$) at corresponding values of the resonant frequency and asymmetry parameter. The geometrical and material parameters of the metasurface are: $p=32$ mm, $h_d=2.5$ mm, $h_s=25.0$ mm, $\varepsilon_d=23.0$, and $\varepsilon_{s}=1.1$.}
\label{fig:simulated}
\end{figure}

In accordance with our available experimental equipment, we have chosen the microwave part of the spectrum to characterize the metasurface and confirm its features. Thus, all the geometrical and material parameters of the dielectric particles as well as the structure period are chosen so that the metasurface operates in the specified frequency band ($1-15$ GHz). All numerical simulations of the electromagnetic response of the metasurface are performed with the use of the rf module of the commercial COMSOL Multiphysics\textsuperscript{\textregistered} electromagnetic solver. The results of simulation of the transmission characteristic of the metasurface versus frequency and asymmetry parameter are collected in Fig.~\ref{fig:simulated} for the three proposed metasurface designs. In these calculations, possible losses in constitutive materials of the metasurface are ignored.

In the spectra of the metasurface composed of non-perturbed disks an additional resonance of reduced transmission appears in the frequency band of interest, provided that the equidistant arrangement of disks in the array is violated [Fig.~\ref{fig:simulated}(a)]. Since this resonance is associated with some violation in the regular array, it can be related to the excitation of the trapped mode. As the asymmetry parameter increases, the quality factor of the resonance decreases and the resonant frequency shifts toward the lower frequencies.

From the cross-section patterns of the electric near-field calculated at the corresponding resonant frequency [see inset of Fig.~\ref{fig:simulated}(a)] one can conclude that the resonant field induced inside each particle resembles the characteristics of the TE$_{01}$ mode of the individual cylindrical dielectric resonator.\cite{Mongia_1994} It arises from the electromagnetic coupling between closely spaced disks in the array with violated periodicity. The electric near-field distribution inside the unit cell exhibits the longitudinal electric dipole moment oriented along the $x$ axis parallel to the direction of the vector $\vec E$ of the incident wave. The magnetic moment induced in each particle is oriented parallel to the $z$ axis and changes sign along the $y$ axis from one particle to another. The resulting magnetic field is concentrated inside the particles, while the electric near-field is distributed in-plane of the metasurface. Electric field is partially concentrated outside the disks, and is localized mainly in the center of cluster. 

Intuition suggests that, the disks should be spaced equidistantly in the array in order to achieve a more homogeneous electric near-field distribution within the cluster. In this case, the trapped mode can be excited by perturbing directly the disks. Indeed, the cluster-based metasurfaces composed of asymmetrically perturbed disks also support the excitation of the TE$_{01}$ mode of the cylindrical dielectric resonator which behaves as a trapped mode of the array.\cite{Yu_JApplPhys_2019, Sayanskiy_PhysRevB_2019} The transmission characteristic for corresponding designs is presented in Figs.~\ref{fig:simulated}(b) and \ref{fig:simulated}(c) for the metasurfaces whose disks are perturbed by the off-centered round hole and coaxial-sector notch, respectively.

For these two designs, the resonance in transmission arises as soon as the perturbation in the disks is introduced and is typical for the trapped mode excitation. The quality factor of the resonance decreases and the resonant frequency shifts toward higher frequencies as the asymmetry parameter increases. The quality factor of the resonance degrades faster for the structure composed of disks with notch. This effect can be explained by vastly reduced extent of dielectric in the cluster, when the notch size increases. 

From the electric near-field patterns plotted in-plane of the metasurface [see insets in Figs.~\ref{fig:simulated}(b) and \ref{fig:simulated}(c)] one can conclude that at the resonant frequency of the trapped mode excitation there is a longitudinal electric dipole moment and magnetic moments, which are induced in a similar way as those of above-discussed array of non-perturbed disks. However, for these two designs, the resulting electric near-field appears to be more homogeneously distributed in-plane of the metasurface and is mainly concentrated outside the disks. The local magnetic field is sufficiently enhanced inside the particles holes and notches. It is important to note, that such near-fields configuration persists even when the asymmetry parameter becomes sufficiently large.

\begin{figure*}[t!]
\centering
\includegraphics[width=1\linewidth]{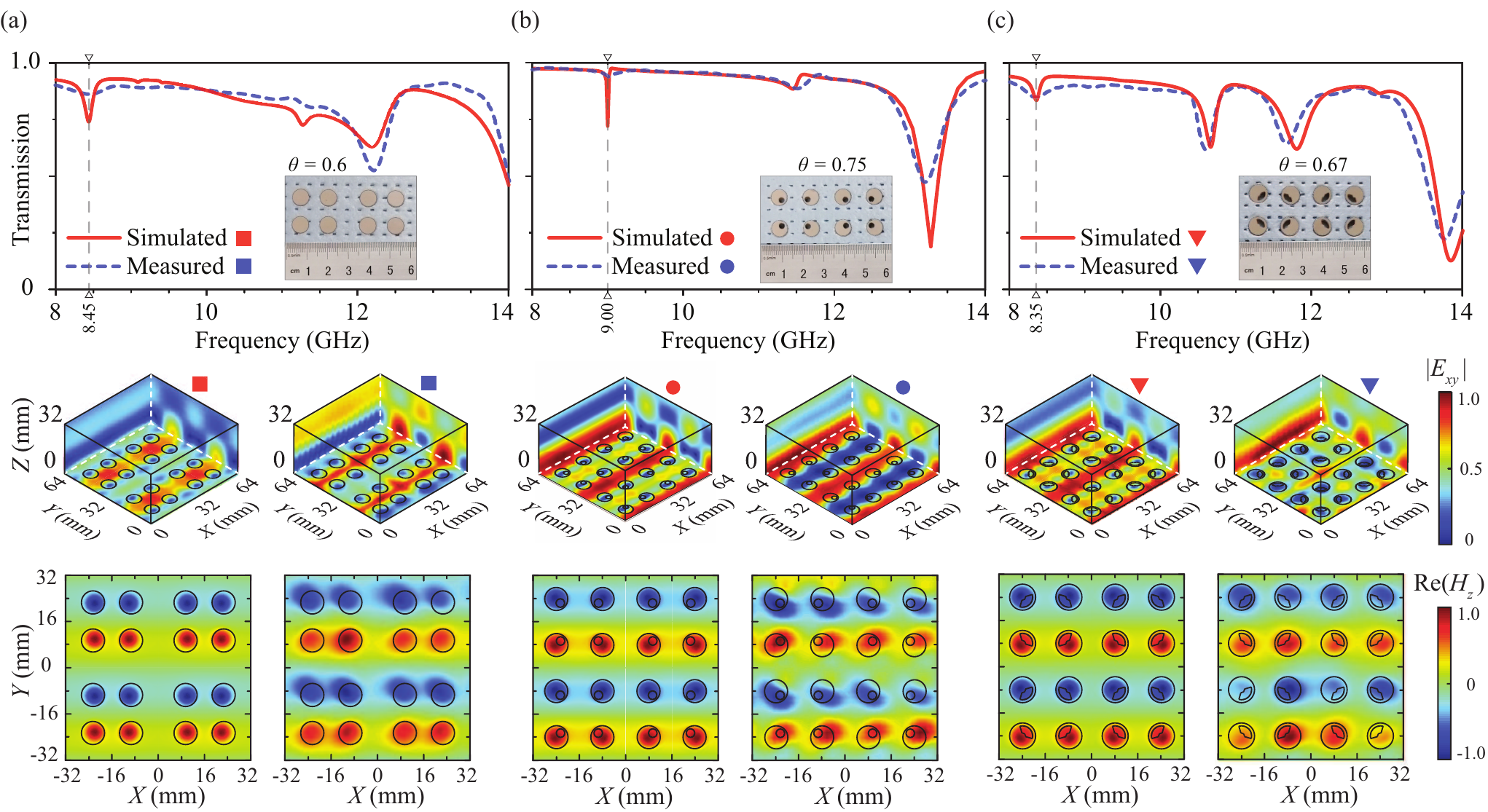}
\caption{Comparison of simulated and measured characteristics of the actual (lossy) cluster-based all-dielectric metasurface. Insets present fragments of the metasurface prototypes. The geometrical and material parameters of the metasurface as well as corresponding values of the resonant frequency and asymmetry parameter coincide with those listed in the caption of Fig.~\ref{fig:simulated}. In the color maps the electric and magnetic near-field magnitudes are normalized on their corresponding maximal values.}
\label{fig:measured}
\end{figure*}  

We check experimentally all the distinctive features of the near-field distribution and the degree of field concentration by direct measurements for the metasurface prototypes based on three chosen designs. The Taizhou Wangling TP-series microwave ceramic is used as a material for particles fabrication (the losses in this material are estimated as $\tan \delta = 1 \times 10^{-3}$ at 10 GHz). The sets of particles are prepared with the use of a precise milling machine. Particles are arranged in a custom holder made of a rigid foam material. All geometrical and material parameters of particles and holder correspond to those listed in the caption of Fig.~\ref{fig:simulated}. All details on the experimental method as well as the sketch and photo of our measurement setup can be found in Refs.~\onlinecite{Sayanskiy_PhysRevB_2019, tuz_AdvOptMat_2019}.

The measured transmission for all proposed designs is depicted in Fig.~\ref{fig:measured} and compared with the results of our numerical simulation, which takes into account estimated material losses in the actual metasurface. The corresponding resonances related to the trapped mode excitation are well identified in the spectra, and they have different quality factors depending on the metasurface design, as predicted above.

We use a near-field scanning system for mapping both electric and magnetic near-fields at the corresponding resonant frequencies. These fields are measured above the metasurface prototypes starting from 1 mm distance from their surface performing subsequent probe moving with the increment of 1 mm in vertical and horizontal directions. The resulting color maps of the measured near-field distribution confirm discussed above trapped mode resonant conditions, distribution of the electric near-field, orientations of the out-of-plane magnetic moments, and concentration of the magnetic near-field inside the particles holes and notches (Fig.~\ref{fig:measured}). In these color maps one can see electric near-field enhancement as a function of the distance to the metasurface array plane. At the resonant frequency, the maximal magnitude of the electric near-field is $1.3\times 10^4$ and $2.0\times 10^4$ times that of the incident field for the metasurfaces composed of non-perturbed and perturbed disks, respectively. Thus, the use of perturbed disks is more advantageous for a stronger electric near-field confinement and furthermore provides more homogeneous field distribution. 

In conclusion, we demonstrate that in the cluster-based metasurface made of an array of equidistantly spaced perturbed high-$n$ dielectric disks, the trapped mode can be excited. For this mode the electric near-field appears to be sufficiently homogeneous and strongly concentrated outside the disks. This effect is extremely promising for applied optics especially for advantageous use of proposed metasurfaces in conjunction with gain media.

V.R.T. acknowledges the hospitality and support of the Jilin University. A.B.E. acknowledges the support of Deutsche Forschungsgemeinschaft (DFG, German Research Foundation) under Germany's Excellence Strategy within the Cluster of Excellence PhoenixD (EXC 2122, Project ID 390833453).

A.S.K. and K.L.D. contributed equally to this work.

\bibliography{spaser.bib}

\end{document}